\documentclass[aps, 11pt, letterpaper, twoside, preprint numbers]{revtex4-1}
\usepackage{amsmath}
\usepackage{hyperref}
\usepackage{amssymb}
\usepackage{amsfonts}

\usepackage[english]{babel}
\usepackage{graphicx}
\begin{document}
\title{$\Large$ The Kelvin Water Dropper: Converting a physics toy into an educational device}
\author{Shreyash Garg$^1$, Rahul Shastri$^2$, B. R. Sivasankaran$^3$}
\author{Luxmi Rani$^4$}\email{luxmi@prl.res.in}
\author{ Bipin K Kaila$^4$}
\author{Navinder Singh$^4$}\email{navinder@prl.res.in}

\affiliation{$^1$S.V. National Institute of Technology, Surat}
\affiliation{$^2$Indian Institute of Technology, Gandhinagar}
\affiliation{$^3$Madras Christian College, Chennai}
\affiliation{$^4$Physical Research Laboratory, Ahmedabad}

\begin{abstract}
The Kelvin Water Dropper was discovered by Lord Kelvin in 1867 and it works on the principle of electrostatic induction. A working model of Kelvin Water Dropper is fabricated in the workshop facility of Physical Research Laboratory, Ahmedabad. We, for the first time, performed a quantitative measurement of the temporal development of charge using a new method that we call ``Effective Capacitance method". With this, the Kelvin Water Dropper experiment can be introduced in undergraduate curriculum, where students can perform quantitative measurements with the apparatus using our ``Effective Capacitance method". This should change the generally held view of Kelvin water dropper as being an entertaining toy to a mature educational device.
\end{abstract}

\maketitle

\section{Introduction}
\noindent
The Kelvin Water Dropper (KWD) also known as Kelvin hydroelectric generator, Kelvin electrostatic generator or Lord Kelvin's thunderstorm was invented in 1867 by Scottish scientist William Thomson (Lord Kelvin) \cite{lord}. It works on the principle of electrostatic induction \cite{knapen,shawn,se,paul,lester,markus,micro}. The details of the charging mechanism are explained in the following sections. In the next section we introduce the basic components of setup.
\section{Experimental setup : Apparatus details}
The experimental setup which we fabricated, and the schematic of Kelvin Dropper are shown in the figure 1 and figure 2 respectively. The experiment setup consists of a water reservoir at a height from the ground which is kept above the apparatus (the reservoir is not shown in the figures). The water from this reservoir is allowed to flow through a pipe whose flow rate can be controlled with the help of a stopper. The water flowing through the glass tube is then split into two stream as shown in figure 2. Each stream of water passes through hollow cylindrical copper tubes before falling into aluminum cans, in which the water is collected. The aluminium cans and copper cylinders are cross connected using copper wires. These connecting wires are attached to crocodile clips in order to make the connections secure. The last component of kelvin water dropper setup is the spark plugs, We use two set of spark plugs, one spherical and another cylindrical, with a conical end (as shown in figures 3 and 4 respectively). We then connect one spark plug (either cylindrical or spherical) to each container. The separation of the spark plug is variable and it is one of the central parameter in our experiment. The containers and the spark plugs are placed on a plexiglass plate in order to insulate the system.
\begin{figure}[htp]
\centering
\includegraphics[width=10cm]{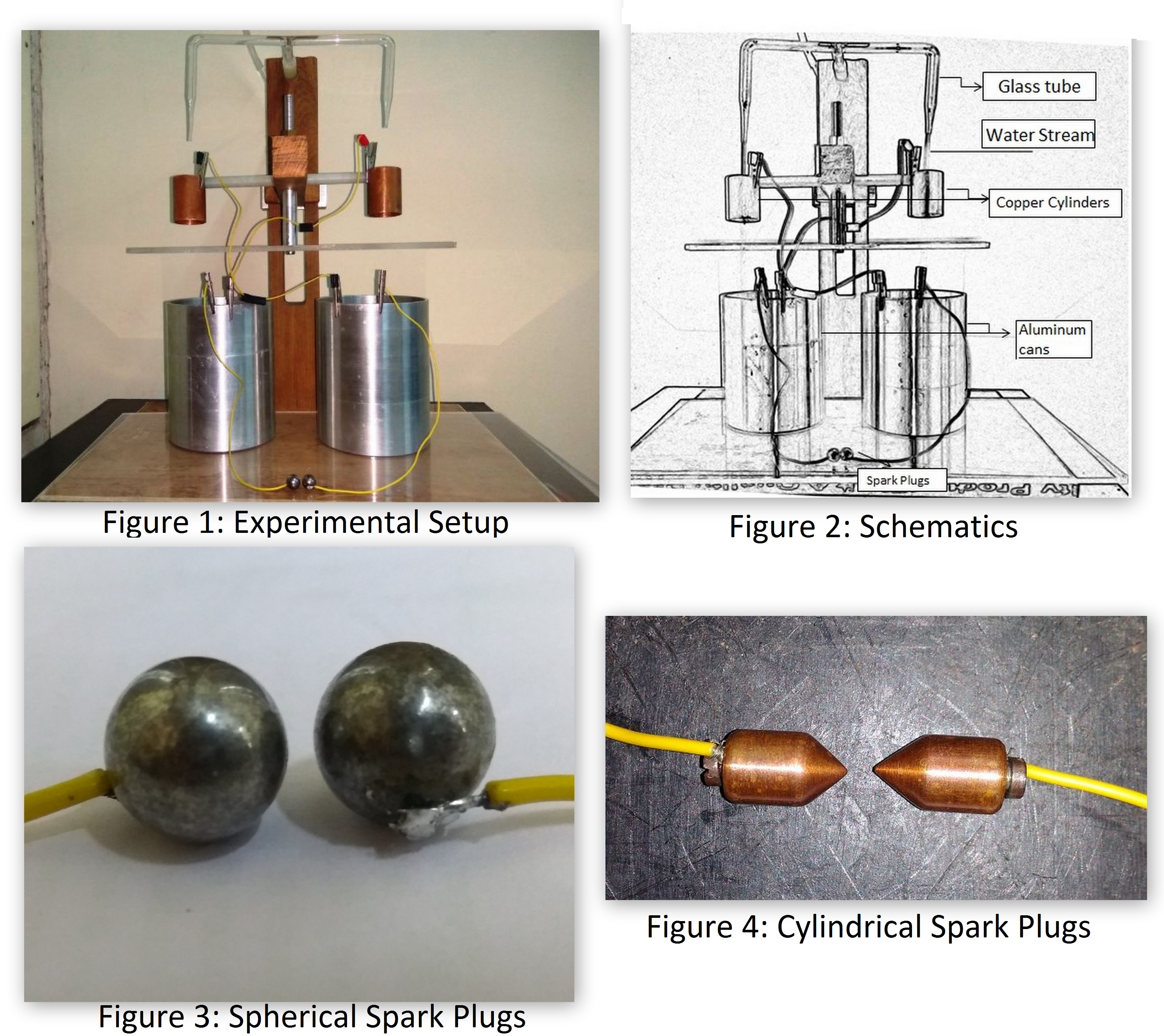}
\end{figure}
The apparatus has two aluminum cylindrical containers of length 159.2 mm and diameter 126.5 mm. It contains two copper cylinders (inductors) of length 50.5 mm and diameter 38.1 mm. Spark plug consists of two spheres having diameter of 12.66 mm and two cylindrical ones with cylindrical height of 15 mm, conical height of 6 mm and cylindrical diameter of 10 mm. The cone angle is $\approx$  $80^o$.

The apparatus is set into operation by opening the stopper and a steady flow is maintained. Water streams pass through copper cylinders and then it is collected into aluminium cans kept beneath the nozzle openings (Figure 1). Within minutes of starting the water flow, one observes electric sparks between the spark plugs. These occur on roughly regular intervals of time. For example, in our apparatus with 1 mm gap between the spark plugs, sparks occurred at average time interval of 2 seconds.
\section{Working Principle}
The working principle of Kelvin Water Dropper is based on electrostatic induction. The induction mechanism can be understood in the following way. To initiate charging process we need seed charge i.e., charge imbalance at the start of experiment.
\setcounter{figure}{4}
\begin{figure}[htp]
\centering
\includegraphics[width=6cm]{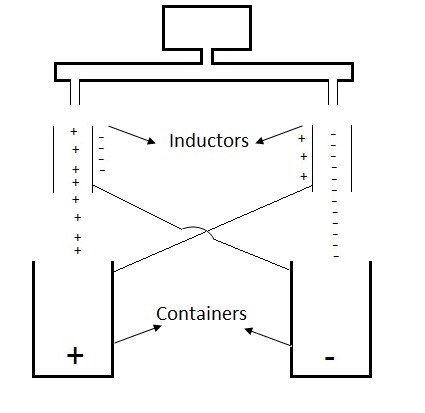}
\caption{Operating principle}
\end{figure}
The seed charge is formed because of collision of air molecules with falling water droplets. Air molecules possess enough kinetic energy to knock some electrons out of water molecules making water stream slightly positively charged. This is exactly what happens in clouds where droplets get charged due to collisions facilitated by wind. If we consider that the stream on left, as shown in figure 5, has momentarily positive charge due to collision with air molecules, the water accumulating in left aluminium can will make the can positively charged. This charge is then transferred to its corresponding copper cylinder connected to it on the right side making it positively charged as well. Electric field of this cylinder attracts a small fraction of $OH^-$ ions or negative charge in falling water which are generally present due to self-ionization of water (self-ionization of water has significant value at room temperature) thus giving us a negatively charged stream on right side which falls into right aluminum can \cite{desmet}. This charge is then transferred to left cylinder which now attracts a small fraction of $H^+$ ions or positive charge. This completes a positive feedback loop, creating a self-amplifying process until voltage reaches breakdown voltage of air (Breakdown voltage of air is 3 kV/mm). It is to be noted that water falling in the right can will have very tiny fraction of $OH^-$ ions and water falling in the left can will have very tiny $H^+$ ions (Refer Appendix A). Thus the chemical imbalance produced will be tiny. As both the cans are connected to spark plugs, there is accumulation of positive charge on left spark plug and negative charge on right spark plug. Once electric field is strong enough we can see a fairly bright spark as discharge occurs. Depending on the distance between spark plugs, there are sparks every few seconds continuously.
Another way to understand formation of seed charge is via Johnson-Nyquist noise which is the electronic noise generated by thermal agitation of the charge carriers i.e. electrons inside an electrical conductor at equilibrium \cite{jn}. We can consider our aluminium and copper cylinders as two conductors connected to each other. Johnson and Nyquist formulated that due to thermal agitation, an electromotive force is generated causing a small current to be present in the circuit. This can cause excess of electrons in copper cylinder temporarily. But this mechanism is ruled out because the fluctuations are at nanoseconds scale which is not sufficient time to sustain charging process.
\section{Measuring temporal development of charge on spheres}
The apparatus is an interesting and entertaining physics toy. However, a deeper understanding of the charging mechanism can be obtained by quantitative measurements. In a typical experiment we measured the time taken for ten sparks for a given separation between the spark plugs. Tables given below display our measurements taken at three different runs of the experiment under the identical environment conditions. We measured the average time taken between the sparks as a function of separation between plugs. The accumulated data is plotted in figure 6 and figure 7. For curve fitting, MATLAB's polyfit function was utilized.
\begin{center}
\begin{tabular}{| c   c   c|} 
\hline
Sr.no & Distance between plugs(mm) & Average time per spark \\ 
\hline
 1 & 1.0 & 2.1 \\ 
\hline
 2 & 2.5 & 2.39 \\ 
\hline
 3 & 4.1 & 3.3 \\ 
\hline
 4 & 7.0 & 6.16 \\ 
\hline
\end{tabular}
\end{center}
\begin{center}
Table 1: Distance between the plugs vs. average time
\end{center}
\begin{center}
\begin{tabular}{| c c c|} 
\hline
Sr.no & Distance between plugs(mm) & Average time per spark \\ 
\hline
 1 & 1.0 & 3.25 \\
\hline
 2 & 3.0 & 6.14 \\ 
\hline
 3 & 6.0 & 7.4 \\ 
 \hline
\end{tabular}
\end{center}
\begin{center}
Table 2: Distance between the plugs vs. average time
\end{center}
\begin{center}
\begin{tabular}{| c c c|} 
\hline
Sr.no & Distance between plugs(mm) & Average time per spark \\ 
\hline
 1 & 0.6 & 2.26 \\ 
\hline
 2 & 1.33 & 2.62 \\ 
\hline
 3 & 2.0 & 2.54 \\
\hline
 4 & 3.4 & 2.65 \\
 \hline
\end{tabular}
\end{center}
\begin{center}
Table 3: Distance between the plugs vs. average time
\end{center}
\begin{figure}[htp]
\centering
\includegraphics[width=10cm]{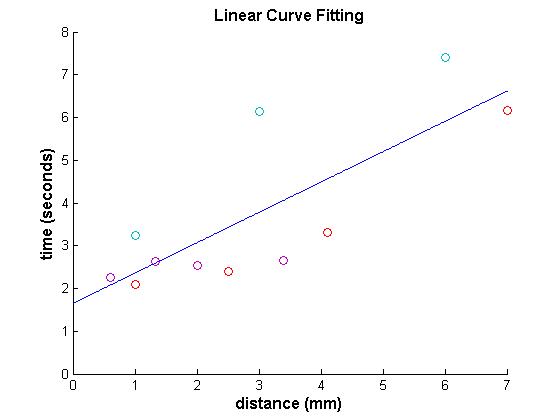}
\caption{Linear Curve Fitting. Straight line is the fitted curve while scattered points are data}
\label{}
\end{figure}
The best fit in the linear case is given by:
\begin{equation}
    y = 0.7095x + 1.6504.
\end{equation}
Here, x is the distance between plugs in mm and y is the average time in seconds.
\newpage
\begin{figure}[htp]
\centering
\includegraphics[width=10cm]{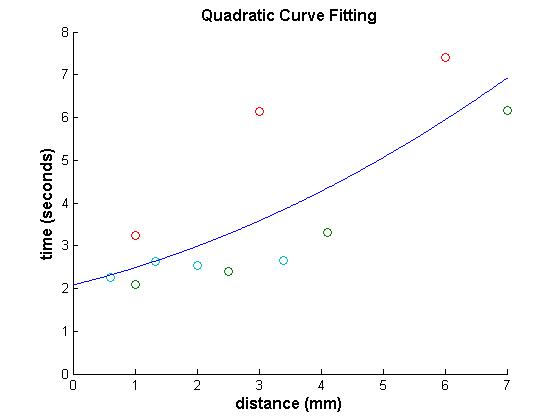}
\caption{Quadratic Curve Fitting. Curved line is the fitted curve while scattered points are data}
\label{}
\end{figure}
The best fit for the quadratic equation is given by:
\begin{equation}
    y = 0.0481x^2 + 0.3546x + 2.0833.
\end{equation}
To investigate which behavior (linear or quadratic) is a faithful representation of data, we calculated the value of 
\begin{equation}
    \sum_{i=1}^{n} (y_i - y^d_i)^2  
\end{equation}
where y$_i$ is a point on the curve (linear or quadratic) and y$^d_i$ is the corresponding data point for the same value of abscissa. We observed that quadratic fitting gives the better representation.
\section{Effective capacitance method and discussion}
We can assume our system of two spark plugs (spherical or conical) as a capacitor with air as dielectric medium between them. The total capacitance of setup is made up of capacitance of aluminum can $C_{can}$, connecting wires $C_{wire}$ and spark plugs $C_{sp}$. These three components are connected in parallel whose effective capacitance can be calculated by adding individual capacitance
\begin{equation}
C_{effective}=C_{can} + C_{wire} + C_{sp}    
\end{equation}
But the capacitance of cans and wires is negligible in comparison to that of spark plugs as the plugs are separated by much smaller distance, Thus
\begin{equation}
C_{effective}\approx C_{sp}    
\end{equation}
\begin{equation}
Q_{ac}=C_{sp}V
\end{equation}
\begin{equation}
Q_{ac}=C_{sp}E_{air}d. 
\end{equation}
Where $E_{air}$ is the breakdown electric field of air at which the sparks are observed and d is the distance between the plugs. It turns out that the capacitance is a weak function of d in comparison to linear one as shown in appendix B. Thus, measuring d vs. average time between the sparks is equivalent to measuring the development of charge ($Q_{ac}$ $\propto$ d) vs. average time. In other words figure 6 and 7 represent the temporal development of charge. We have observed that the quadratic fitting represent the phenomenon in a better way. We can also infer this fact from the positive feedback mechanism discussed in section 3. The rate of charge accumulation should enhance as more positive or negative charge getting accumulated on copper cylinders further polarizes water and leads to enhanced charge accumulation rate in the aluminum cans. This leads to an amplifying feedback as these are cross-connected to copper cylinders. Ultimately charge accumulation should lead to saturation due to: (a) Spraying effect of water, and (b) the internal Coulomb barrier. The “spraying effect of water" is observed at larger separation between the plugs. This leads to larger charge accumulation. As water droplets of positive charge (for example) fall into positively charged aluminum can, they feel electrostatic repulsion (because they have same charge). When the energy gain due to fall under earth gravity is less than the increase in energy due to electrostatic repulsion, it is no more energetically favorable for the drops to fall in the container. Droplets start to fall outside and are not collected in the container. This spraying or spillage of water leads to charge leakage and saturation happens. The second factor of “internal Coulomb barrier" also occurs in the water itself when it separates inside the glass tubes. The oppositely charge water will have electrostatic attraction and it will retard the motion to water into two separate nozzles. These factors both combined lead to saturation of the charge accumulation. In our experiment it is difficult to take readings with large separation due to spraying effect. Water spill over leads to leakage and the apparatus stops functioning.

\section{Appendix A}
The excess of $H^+$ and $OH^-$ ions will be negligible in comparison to amount of $H_2O$ molecules present. For this, let's assume the aluminum can has at any instant, 1 liter of water inside of it. Thus moles of water molecules assuming density of water as 1gm/cc are
\begin{equation}
    1000/18.015 \approx 55.5\: moles.
\end{equation}
Giving us amount of water molecules as 
\begin{equation}
    55.5*N_A=55.5*6.023*10^{23}=3.3*10^{25} \:molecules.
\end{equation}
Amount of ions can be calculated by formulating approximate value of charge developed on spheres. As the breakdown voltage is 3kV/mm, voltage across plugs is of order $10^3$ V and capacitance is of the order $10^{-12}$ F. Thus charge developed is around $10^{-9}$ C.
So amount of ions is
\begin{equation}
    \approx \frac{10^{-9}}{10^{-19}}=10^{10} \:ions.
\end{equation}
Thus, fraction of ions accumulated comes out to be
\begin{equation}
    \approx \frac{10^{10}}{10^{25}}=10^{-15}.
\end{equation}
We can thus conclude that chemical imbalance due to ions will be negligible.
\section{Appendix B}
We can calculate the maximum amount of charge accumulated on spheres just before the sparks. For this, we need to calculate capacitive coefficient for two sphere system first of all. The formal solution of the capacitive coefficient is in the form of infinite series slowly converging when two spheres are too close to each other. The problem can be reduced to a simple case of a conducting sphere interacting with a conducting wall. An approximate simple expression for the capacitive coefficient for the sphere and wall problem was derived by Crowley \cite{crowley}. Capacitance of the two spheres separated by a distance is given by the expression
\begin{equation}
    C_{crowley}=4\pi \epsilon a[1 + 0.5 log(1+\frac{2a}{r-2a} ) ].
\end{equation}
Here a is the radius of the sphere and r is the distance between sphere's centers. So
\begin{equation}
    Q_{crowley}=C_{crowley}V=C E_{air}(r-2a)
\end{equation}
\begin{equation}
  Q_{crowley}  =4\pi \epsilon a[1 + 0.5 log(1+\frac{2a}{r-2a} ) ]E_{air}(r-2a)
\end{equation}
\begin{equation}
  Q_{crowley}   =4\pi \epsilon a E_{air}[1 + 0.5 log(1+\frac{2a}{d} ) ]d.
\end{equation}
Here
\begin{equation}
     d=r-2a
\end{equation}
d is the width of gap between the spheres.

In figures 8(a) and 8(b) we plot $Q_{crowley}$ vs. d and $C_{crowley}$ vs. d/a. k here is an arbitrary constant. It is clear that $Q_{crowley}$ is roughly linear in d (logarithmic functions are weaker functions).
\begin{figure}[htp]
\centering
\includegraphics[width=12cm]{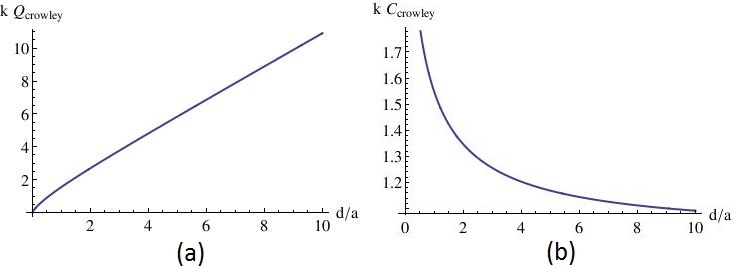}
\caption{(a) Plot of $Q_{crowley}$ vs. d. and (b) $C_{crowley}$ vs. d/a.}
\label{}
\end{figure}

\end{document}